\documentclass[12pt,reqno]{article}
\usepackage{amsmath}
\usepackage{graphicx}
\usepackage{color}
\usepackage{varioref}

\begin{document}

\title
\centerline{\bf\large{Nonlinear GLR-MQ evolution equation and $Q^2$-evolution}}
\centerline{\bf\large{of gluon distribution function}}
\vspace{7pt}
\vspace{7pt}

\centerline{{\bf Mayuri Devee}$^{\rm *}$ and {\bf J. K. Sarma}}
\centerline{\small{HEP Laboratory, Department of Physics, Tezpur University, Napaam 784 028, Tezpur, Assam,  India}} 
\centerline{$^{\rm *}${\it deveemayuri@gmail.com}}

\vspace{7pt}
\vspace{7pt}

\renewcommand{\baselinestretch}{1.50}\normalsize

\begin{abstract}
\leftskip0.1mm
\rightskip0.1mm
 In this paper we have solved the nonlinear Gribov-Levin-Ryskin-Mueller-Qiu (GLR-MQ) evolution equation for gluon distribution function $G(x,Q^2)$  and studied the effects of the nonlinear GLR-MQ corrections to the Leading Order (LO) Dokshitzer-Gribov-Lipatov-Altarelli-Parisi (DGLAP) evolution equations. Here we incorporate a Regge like behaviour of gluon distribution function to obtain the solution of GLR-MQ evolution equation. We have also investigated the $Q^2$-dependence of gluon distribution function from the solution of GLR-MQ evolution equation. Moreover it is interesting to observe from our results that nonlinearities increase with decreasing correlation radius ($R$) between two interacting gluons. Results also confirm that the steep behavior of gluon distribution function is observed at $R=5\hspace{4pt}GeV^{-1}$, whereas  it is lowered at $R=2\hspace{4pt}GeV^{-1}$ with decreasing $x$ as $Q^2$ increases. In this work we have also checked the sensitivity of $\lambda_G$ in our calculations. Our computed results are compared with those obtained by the global DGLAP fits to the parton distribution functions viz. GRV, MRST, MSTW and with the EHKQS model.

\ Keywords: {GLR-MQ equation, gluon distribution function, DGLAP equation}

\ PACS no. {12.38.-$t$, 12.39.-$x$, 12.38.-$Bx$, 13.60.-$Hb$, 13.85.-$Hd$}

\end{abstract}

\section{Introduction}
\paragraph\ The small-$x$, where $x$ is the Bjorken scaling variable, behavior of quark and gluon densities is one of the challenging problems of quantum chromodynamics (QCD). The most important phenomena in the region of small-$x$  which determine the physical picture of the parton (quark and gluon) evolution or cascade, are the increase of the parton density at $x\rightarrow{0}$, the growth of the mean transverse momentum of a parton inside the parton cascade at small-$x$, and the saturation of the parton density [1]. The parton  distributions in hadrons play a key role in understanding the standard model processes and in the predictions for such processes at accelerators. 
Therefore the determination of parton densities or more importantly the gluon  density in the small-$x$ region is particularly interesting because here gluons are expected to dominate the proton structure function. The study of gluon distribution function is also very important because it is the basic ingredient in the calculations of different high-energy hadronic processes like mini jet production, growth of total hadronic processes etc. Moreover precise knowledge of the gluon distribution at small-$x$ is essential for reliable predictions of important p-p, p-A and A-A processes studied at the relativistic heavy-ion collider (RHIC)[2] and at CERN$^\prime$s large hadron collider (LHC) [3]. Knowledge of gluon density is also important for the computation of inclusive cross-sections of hard, collinearly factorizable, processes in hadronic collisions.

\ The most precise determinations of the gluon momentum distribution in the proton can be obtained from a  measurement of the deep inelastic scattering (DIS) proton structure function $F_2(x,Q^2)$ and its scaling violation. The measurement of the proton structure function $F_2(x,Q^2)$ by  H1 [4] and ZEUS [5] at HERA over a broad kinematic region has made it possible to know about the gluon in the formerly unexplored region of $x$ and  $Q^2$ where, $Q^2$ is the virtuality of the exchanged virtual photon. This method is however indirect because $F_2(x, Q^2)$ at low values of $x$ actually probes the sea quark distributions which are related via the QCD evolution equations to the gluon distribution. More direct determinations of the gluon distribution can be obtained by reconstruction of the kinematics of the interacting partons from the measurement of the hadronic final state in gluon induced processes. They are subject to different systematic effects and provide an substantive test of perturbative QCD. Direct gluon density determinations have been carried out using events with $J/\psi$ mesons in the final state [6] and dijet events [7].

\ In perturbative QCD, the high-$Q^2$ behavior of DIS is given by the linear Dokshitzer-Gribov-Lipatov-Altarelli-Parisi (DGLAP) evolution equations [8]. The number density of gluons, $G(x,Q^2)$, and quarks, $q(x,Q^2)$, in a hadron can be evaluated at large-$Q^2$ by solving the linear DGLAP equation to calculate the emission of additional quarks and gluons compared to some given initial distributions. The results are adjusted to fit the experimental data (mainly at small-$x$) for the proton structure function $F_2(x,Q^2)$ measured in DIS, over a large domain of values of $x$ and $Q^2$ by adjusting the parameters in the initial parton distributions.  Consecuently, the approximate analytical solutions of DGLAP evolution equations have been reported in recent years with significant phenomenological success [9-11].

\ DGLAP equation predicts a sharp growth of the gluon distribution function as $x$ grows smaller which is also clearly observed in DIS experiments at HERA. This sharp growth of the gluon distribution function will have to eventually slow down in order to not violate unitarity bound [12] on physical cross sections. It is a known fact that the hadronic cross sections comply with the Froissart bound [12] which derives from the general assumptions of the analyticity and unitarity of the scattering amplitude. The Froissart bound indicates that the total cross section does not grow faster than the logarithm squared of the energy i.e., $\sigma_{total}=\frac{\pi}{m_{\pi}^2}(ln s)^2$, where, $m_{\pi}$ is the scale of the range of the strong force [13]. Gluon recombination is commonly believed to provide the mechanism responsible for the unitarization of the cross section at high energies or a possible saturation of the gluon distribution function at small-$x$. In other words, the number of gluons at small-$x$ will be so large that they will spatially overlap and therefore, gluon recombination will be as important as gluon splitting. In the derivation of the linear DGLAP equation the correlations among the initial gluons in the physical process of interaction and recombination of gluons are usually omitted. But at small-$x$ the corrections of the correlations among initial gluons to the evolutionary amplitude should be taken into account. These multiple gluon interactions induce nonlinear corrections in the DGLAP equation and so the standard linear DGLAP evolution equation will have to be modified in order to take this into effect.

\ The proton structure function $F_2(x,Q^2)$ has been measured  down to $x\sim{10^{-5}}$ but still in the perturbatively accessible region by the H1 Collaboration at HERA [4]. These data have been included in the recent global analyzes by the MRST [14] and CTEQ [15] collaborations. DGLAP evolution equations can describe the available experimental data quite well in a fairly broad range of $x$ and $Q^2$ with appropriate parameterizations. But DGLAP approach cannot provide a good description while trying to fit the H1 data simultaneously in the region of large-$Q^2$ ($Q^2> 4\hspace{4pt}GeV^2$) and in the region of small-$Q^2$ ($1.5\hspace{4pt}GeV^2<Q^2<4\hspace{4pt}GeV^2$) [4, 16]. 
This implies that towards smaller values of $x$ and (or) $Q^2$ (but still $Q^2\geq{\Lambda^2}$, $\Lambda$ being the QCD cut off papameter) it is possible to observe gluon recombination effects which lead to nonlinear power corrections to the DGLAP equations. These nonlinear terms lower the growth of the gluon distribution in this kinematic region where $\alpha_S$ is still small but the density of partons becomes very large. Therefore, the corrections of the higher order QCD effects, which suppress or shadow the growth of parton densities, become a center of intensive study in the last few years.

\ Gribov, Levin, Ryskin, Mueller and Qiu (GLRMQ) performed a detailed study of this region in their pioneering papers and they suggested that these shadowing corrections could be expressed in a new evolution equation known as the GLR-MQ equation [17-18]. This equation, involves a new quantity, $G^2(x,Q^2)$, the two-gluon distribution per unit area of the hadron. The main features of this equation are that it predicts a saturation of the gluon distribution at very small-$x$, it predicts a critical line separating the perturbative regime from the saturation regime and it is only valid in the border of this critical line [16, 19]. It is an amazing property of GLR-MQ equation is that it introduces a characteristic momentum scale $Q_s^2$, which is a measure of the density of the saturated gluons. It grows rapidly with energy, and it is proportional to $1/x^\lambda$ with $\lambda=0.2$ [20]. Gribov, Levin and Ryskin first suggest a nonlinear evolution equation, in which the evolution kernels, which they called as the gluon recombination functions, are constructed by the fan diagrams [17]. Later Mueller and Qiu  calculated the gluon recombination functions at the double leading logarithmic approximation (DLLA) in a covariant perturbation framework [18].

\ The GLR-MQ equation is broadly regarded as a key link from perturbation region to non-perturbation region. There has been much work inspired by the approach of GLR-MQ which show that gluon recombination leads to saturation of gluon density at small-$x$ [21-22]. The predictions of the GLR-MQ equation for the gluon saturation scale were studied in Ref. [16]. A new evolution equation named as modified DGLAP equation is derived by Zhu and Ruan [23] where the applications of the AGK  (Abramovsky-Gribov-Kancheli) cutting rule [24] in the GLR-MQ equation was argued in a more general consideration. Here the Feynman diagrams are summed in a quantum field theory framework instead of the AGK cutting rule. In Ref. [25] parton distribution functions in the small-$x$ region are numerically predicted by using a modified DGLAP equation with the GRV-like input distributions.
Moreover, some studies of the GLR-MQ terms in the framework of extracting the PDFs of the free proton can be found in Ref. [26]. Also other nonlinear evolution equations relevant at high gluon densities have been derived in the recent years, and the structure functions from DIS have been analyzed in the context of saturation models [27-29].  

\ The solution of the GLR-MQ equation is particularly important for understanding the nonlinear effects of gluon-gluon fusion due to the high gluon density at small enough $x$. The solution of nonlinear evolution equations also provides the determination of the saturation momentum that incorporates physics in addition to that of the linear evolution equations commonly used to fit DIS data. Various studies on the solutions and viable generalizations of the GLR-MQ equation have been done in great detail in the last few years [30-32]. In the present work we intend to obtain a solution of the nonlinear GLR-MQ evolution equation for the calculation of gluon distribution function in leading order. This paper addresses interesting questions about validity of the well known Regee like parametrization in the region of moderate virtuality of photon. Here we have also calculated the $Q^2$-evolution of gluon distribution function and the results are compared with the predictions of different paramerizations like GRV1998LO [33], MRST2001LO [14], MSTW2008LO [34] and EHKQS molel [16]. Finally, we present our conclusions.

\section{Theory}
\paragraph\ The GLR-MQ equation is based on two processes in the parton cascade: the emission induced by the QCD vertex $g{\rightarrow}g+g$ with a probability which is proportional to $\alpha_S\rho$ and the annihilation of a gluon by the same vertex $g+g{\rightarrow}g$ with a probability which is proportional to $\alpha_{S}^2{r^2}{\rho}^2$, where $\rho(x,Q^2)$=$\frac{xg(x,Q^2)}{{\pi}R^2}$ is the density of the gluon in the transverse plane, ${\pi}R^2$ is the target area, and $R$ is the correlation radius between two interacting gluons. Normally, this radius should be smaller than the radius of a hadron. It is worthwhile to mention that $R$ is non-perturbative in nature and therefore all physics that happens at distance scales larger than $R$ is non-perturbative [30]. Here, $r$ is the size of the parton (gluon) produced in the annihilation process. For DIS $r{\propto}\frac{1}{Q^2}$. Clearly, at $x\sim1$ only the production of new partons (emission) is essential because $\rho{\ll}1$ , but at $x{\rightarrow}0$ the value of $\rho$ becomes so large that the annihilation of partons becomes important.

\ To take interaction and recombination of partons (mainly gluons) into account, a small parameter is introduced which enables us to estimate the accuracy of the calculation, given as,
\begin{equation}W=\frac{\alpha_s}{Q^2}\rho(x,Q^2),\end{equation}
which is the probability of a gluon recombination during the cascade. Here the first factor ${\alpha_s}/{Q^2}$ is the cross section for absorption of a gluon by a parton in the hadron. The unitarity constraint defined in the introduction can be rewritten in the form $W\leq{1}$ [30]. Thus the amplitudes that include gluon recombination can be represented by a perturbation series in this parameter.

\ The number of partons in a phase space cell ($\Delta{ln(1/x)}\Delta{lnQ^2}$) increases through emission and decreases through annihilation and as a result the balance equation for emission and annihilation of partons can be written as [1, 17-18]

\begin{equation}\frac{\partial^2{\rho}}{{\partial}{ln(1/x)}{\partial}\ln{{Q}^2}}=\frac{{\alpha_s}N_c}{\pi}\rho-\frac{\alpha_s^2\gamma}{Q^2}\rho^2.\end{equation}
In terms of gluon distribution function this equation can be expressed as
\begin{equation}\frac{\partial^2{xg(x,Q^2)}}{{\partial}{ln(1/x)}{\partial}\ln{{Q}^2}}=\frac{{\alpha_s}N_c}{\pi}xg(x,Q^2)-\frac{\alpha_s^2\gamma}{{\pi}Q^2R^2}[xg(x,Q^2)]^2,\end{equation}
which is named as the GLR-MQ evolution equation. The factor $\gamma$ is found to be $\gamma=\frac{81}{16}$ for $N_c=3$, as calculated by Mueller and Qiu [18].

\ Now to study the $Q^2$-evolution of gluon distribution function, we can rewrite Eq. (3) in a convenient form [31]
\begin{equation}\frac{\partial{G(x,Q^2)}}{\partial{\ln{Q^2}}}=\frac{\partial{G(x,Q^2)}}{\partial{ln{Q^2}}}\Big\vert_{DGLAP}-\frac{81}{16}{\frac{\alpha_S^2(Q^2)}{{R^2}{Q^2}}}\int_x^1\frac{d\omega}{\omega}{G^2\Big(\frac{x}{\omega},Q^2\Big)},\end{equation}
where the first term in the r.h.s. is the usual linear DGLAP term in the double leading logarithmic approximation and the second term is nonlinear in gluon density.



\ Here, the representation for the gluon distribution $G(x,Q^2)=xg(x,Q^2)$ is used, where $g(x,Q^2)$ is the gluon density. The quark gluon emission diagrams are neglected due to their little importance in the gluon-rich small-$x$ region. The negative sign in front of the non-linear term is responsible for the gluon recombination. The strong growth generated by the linear term is lowered by the non-linear term for large gluon densities and so it describes shadowing corrections. The size of the non-linear term depends on the value of $R$.  For $R=R_h$ shadowing corrections is negligibly small whereas, for $R{\ll}R_h$ shadowing corrections is expected to be large, $R_h$ being the radius of the hadron in which gluons are populated [1, 30].

\ To simplify our calculations we consider a variable $t$ such that $t=\ln\frac{Q^2}{\Lambda^2}$, where $\Lambda$ is the QCD cut off parameter. Then Eq. (4) becomes
\begin{equation}\frac{\partial{G(x,t)}}{\partial{t}}=\frac{\partial{G(x,t)}}{\partial{t}}\Big\vert_{DGLAP}-\frac{81}{16}\frac{\alpha_S^2(t)}{{R^2}{\Lambda^2}e^t}\int_x^1\frac{d\omega}{\omega}{\Big[G\Big(\frac{x}{\omega},t\Big)\Big]}^2.\end{equation}

As gluons are the dominant parton at small-$x$, therefore, ignoring the quark contribution to the gluon distribution function the first term in the r.h.s. of Eq. (5) can be expressed as [35]
\begin{displaymath}\frac{\partial{G(x,t)}}{\partial{t}}\Big\vert_{DGLAP}=\alpha_S(t)\Big[\Big(\frac{11}{12}-\frac{N_f}{18}+\ln(1-x)\Big)G(x,t)\end{displaymath}
\begin{equation}+\int_x^1d{\omega}\Big\{\frac{\omega{G\Big(\frac{x}{\omega},t\Big)}-G(x,t)}{1-\omega}+\Big(\omega(1-\omega)+\frac{1-\omega}{\omega}\Big)G\Big(\frac{x}{\omega},t\Big)\Big\}\Big].\end{equation}
The strong coupling constant $\alpha_S{(t)}$ in leading order has the form [35]
\begin{equation}\alpha_{S}{(t)}=\frac{4\pi}{\beta_0{t}},\end{equation}
where, \begin{equation}\beta_{0}=\frac{11}{3}{N_c}-\frac{4}{3}{T_f}=11-\frac{2}{3}{N_f}\end{equation}
is the one-loop corrections to the QCD $\beta$-function and $N_f$ being the number of quark flavor. Here we consider $N_{c}=3$, and ${T_{f}}={{\dfrac{1}{2}}{N_{f}}}$ and $N_f=4$.

\ At small-$x$, the behavior of structure functions is well explained in terms of Regge-like behavior [36, 37]. The small-$x$ behaviour of structure functions for fixed $Q^2$ reflects the high-energy behavior of the total cross section with increasing total CM energy squared $s^2$, since $s^2=Q^2(\frac{1}{x}-1)$ [38]. The Regge pole exchange picture [37] would therefore appear quite appropriate for the theoretical description of this behaviour. The Regge behavior of the sea-quark and antiquark distribution for small-$x$ is given by $q_{sea}(x)\sim{x^{-\alpha_P}}$ corresponding to a pomeron exchange with an intercept of $\alpha_P=1$. But the valence-quark distribution for small $x$ given by $q_{val}(x)\sim{x^{-\alpha_R}}$ corresponds to a reggeon exchange with an intercept of $\alpha_R=0.5$. The $x$ dependence of the parton densities is often assumed at moderate $Q^2$ and thus the leading order calculations in $ln(1/x)$ with fixed $\alpha_S$ predict a steep power-law behavior of $xg(x,Q^2)\sim{x^{-\lambda_G}}$, where $\lambda_G=({3\alpha_S}/{\pi})4ln2\simeq{0.5}$ for $\alpha_S\simeq{0.2}$, as appropriate for $Q^2\sim4 GeV^2$.

\ Moreover the Regge theory provides extremely naive and frugal parameterization of all total cross sections [39, 40]. It is suggested in Refs. [41, 42] that is feasible to use Regge theory for the study of DGLAP evolution equations. The tactics for the determination of the gluon distribution function with the nonlinear correction is also based on the Regge-like behavior [43]. The Regee behavior is believed to be valid at small-$x$ and at some intermediate $Q^2$, where $Q^2$ must be small, but not so small that $\alpha_S(Q^2)$ is too large [44, 45]. Moreover, as discussed in [40] the Regge theory is supposed to be applicable if $W^2$ is much greater than all the other variables and so, models based upon this idea have been successful in describing the DIS cross-section when $x$ is small enough ($x<0.01$), whatever be the value of $Q^2$. [20, 46].

\ Therefore, to solve the GLR-MQ equation, we consider a simple form of Regee like behavior for the determination of the gluon distribution function at small-$x$ given as
\begin{equation}G(x,t)=M(t)x^{-\lambda_G}\end{equation}

which implies, \begin{equation}G\Big(\frac{x}{\omega},t\Big)=M(t)x^{-\lambda_G}\omega^{\lambda_G}=G(x,t)\omega^{\lambda_G},\end{equation}
and \begin{equation}G^2\Big(\frac{x}{\omega},t\Big)=\{M(t)x^{-\lambda_G}\}^2\omega^{2\lambda_G}=G^2(x,t)\omega^{2\lambda_G},\end{equation}
where $M(t)$ is a function of $t$ and $\lambda_G$ is the Regge intercept for gluon distribution function. This form of Regge behaviour is well supported by the work of the authors in Refs. [40, 47, 48]. According to Regge theory, the high energy i.e. small-$x$ behaviour of both gluons and sea quarks are controlled by the same singularity factor in the complex angular momentum plane [37]. Moreover, as the values of Regge intercepts for all the spin-independent singlet, non-singlet and gluon structure functions should be close to 0.5 in quite a broad range of small-$x$ [48], we would also expect that our theoretical results are best fitted to those of the experimental data and parameterization at $\lambda_G\approx{0.5}$, where $\lambda_G$ is the Regge intercepts for gluon distribution function.

\ Substituting Eqs. (6), (10) and (11) in Eq. (5) we get

\begin{displaymath}\frac{\partial{G(x,t)}}{\partial{t}}=\alpha_S(t)G(x,t)\Big[\Big(\frac{11}{12}-\frac{N_f}{18}+\ln(1-x)\Big)+\int_x^1d{\omega}\Big\{\frac{\omega^{\lambda_G+1}-1}{1-\omega}+\end{displaymath}
\begin{equation}\Big(\omega(1-\omega)+\frac{1-\omega}{\omega}\Big)\omega^{\lambda_G}\Big\}\Big]-\frac{81}{16}\frac{\alpha_S^2(t)}{{R^2}{\Lambda^2}e^t}G^2(x,t)\int_x^1\omega^{2\lambda_G-1}d\omega.\end{equation}
Performing the integrations and rearranging the terms, Eq. (12) takes the form
\begin{equation}\frac{\partial{G(x,t)}}{\partial{t}}=P(x)\frac{G(x,t)}{t}-Q(x)\frac{G^2(x,t)}{t^2e^t},\end{equation}
with,
\begin{equation}P(x)=\frac{4\pi}{\beta_0}\Big[\frac{11}{12}-\frac{N_f}{18}+\ln(1-x)+\Big(\frac{2}{2+\lambda_G}+\frac{1}{\lambda_G}-1\Big)-\Big(\frac{2x^{2+\lambda_G}}{2+\lambda_G}+\frac{x^{\lambda_G}}{\lambda_G}-x\Big)\Big]\end{equation}
and\begin{equation}Q(x)=\frac{81\pi^2}{2R^2\Lambda^2\beta_0^2}\Big(\frac{1-x^{2\lambda_G}}{\lambda_G}\Big)\end{equation}
Eq. (13) is a partial differential equation which can be solved as
\begin{equation}G(x,t)=\frac{t^{P(x)}}{C-Q(x)\Gamma[-1+P(x),t]},\end{equation}
where $\Gamma$ is the incomplete gamma function and C is a constant. Although Regge behavior is not in agreement with the double-leading-logarithmic solution, namely, $G(x,t)\propto{exp[C ln(t) ln(1/x)]^{1/2}}$, but, the range where $x$ is small and $Q^2$ is not very large is actually the Regge regime. Accordingly solution of the GLR-MQ equation in the form of Eq. (16) is expected to be worthwhile. We believe that our solution is correct in the vicinity of the saturation scale where all our assumptions look natural.

\ It is clear from Eq. (16) that at large $t$, we can neglect the nonlinear corrections and our solution takes the form
\begin{equation}G(x,t)=\frac{t^{P(x)}}{C-Q(x)\Gamma[-1+P(x),t]}\stackrel{t>>1}{\longrightarrow}{t^P(x)/C},\end{equation}
However, in the region where $t$ is not very large the corrections for the nonlinear term in Eq. (16) can not be neglected and therefore Eq. (16) does not reduce to Eq. (17). Thus we can expect that the solution given by Eq. (16) is only valid in the region of small-$x$ and intermediate values of $Q^2$ (or $t$).

\ Now, to determine the $Q^2$-dependence of $G(x, Q^2)$, we apply initial conditions at $t=t_0$ where, $t_0=\ln\Big(\frac{Q_0^2}{\Lambda^2}\Big)$ for any lower value of $Q=Q_0$, to get
\begin{equation}G(x,t_0)=\frac{t_0^{P(x)}}{C-Q(x)\Gamma[-1+P(x),t_0]},\end{equation}
from which we obtain the value of the constant C as,
\begin{equation}C=\frac{t_0^{P(x)}+Q(x)\Gamma[-1+P(x),t_0]G(x,t_0)}{G(x,t_0)}.\end{equation}
From this equation the constant C can be evaluated by considering an appropriate input distribution $G(x,t_0)$ at a given value of $Q_0^2$. Now substituting C from Eq. (19) in Eq. (16) we obtain the $Q^2$-evolution of gluon distribution function for fixed $x$ in leading order as

\begin{equation}G(x,t)=\frac{t^{P(x)}G(x,t_0)}{{t_0^{P(x)}}+Q(x)\Big\{\Gamma[-1+P(x),t_0]-\Gamma[-1+P(x),t]\Big\}G(x,t_0)}.\end{equation}

\ Thus we have obtained an expression for the $Q^2$-evolution of gluon distribution function $G(x, t)$ in leading order by solving the nonlinear GLR-MQ evolution equation semi-numerically. From the final expression given by Eq. (20) we can easily calculate the $Q^2$-evolution of $G(x, Q^2)$ for a particular value of $x$ by taking an appropiate input distribution at a given value of $Q_0^2$.

\section{Result and discussion}
\paragraph \ In this paper we have solved the nonlinear GLR-MQ evolution equation in order to determine the $Q^2$-dependence of gluon distribution function $G(x,Q^2)$. We have compared our results of $Q^2$-evolution of $G(x,Q^2)$ with those obtained by the global DGLAP fits to the parton distribution functions GRV1998LO [31], MRST2001LO [13], MSTW2008LO [32] respectively. We have also compared our computed results with the EHKQS [16] model. The GRV1998 global parametrization used H1 and ZEUS high precision data on $G(x,Q^2)$. The MRST2001  parametrization is a global analyses of data which include the new precise data on DIS from HERA together with constraints from hard scattering data. MSTW2008 presented an updated parton distribution functions  determined from global analysis of hard-scattering data within the standard framework of leading-twist fixed-order collinear factorisation in the $\overline{MS}$ scheme. These parton distributions supersede the previously available MRST sets and can be used for the first LHC data taking and for the associated theoretical calculations. In EHKQS model [16] the effects of the first nonlinear corrections to the DGLAP evolution equations are studied by using the recent HERA data for the structure function $F_2(x, Q^2)$ of the free proton and the parton distributions from CTEQ5L [14] and CTEQ6L [14] as a baseline. By requiring a good fit to the H1 data, they determine initial parton distributions at $Q_0^2=1.4\hspace{4pt}GeV^2$ for the nonlinear scale evolution. In Ref. [16] it is shown that the nonlinear corrections enhance the agreement with the  $F_2(x, Q^2)$ data in the region of $x\sim$ $3$x$10^{-5}$ and $Q^2\sim{1.5\hspace{4pt}GeV^2}$.

\begin{figure}[!htb]
\centering
\includegraphics[width=2.5in,height=3in]{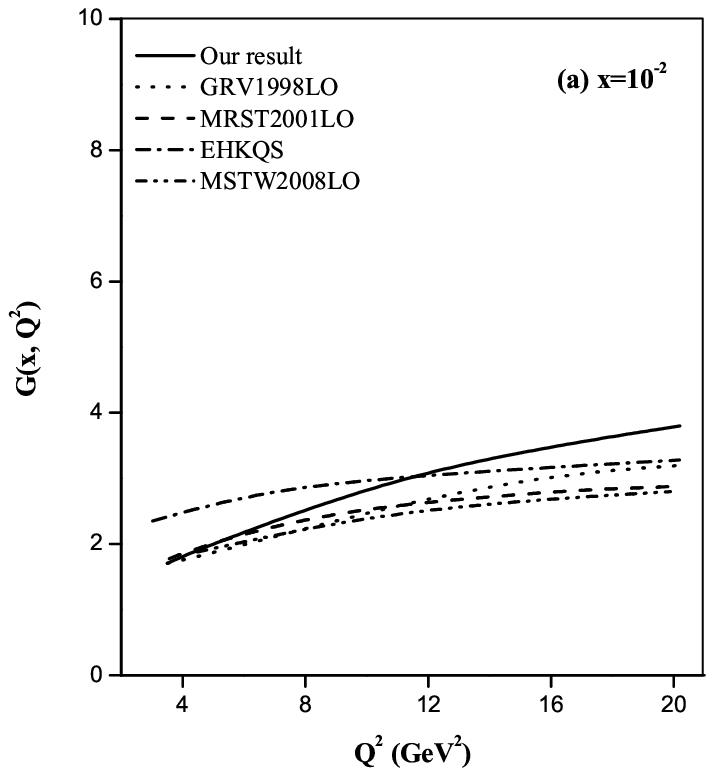}
\includegraphics[width=2.5in,height=3in]{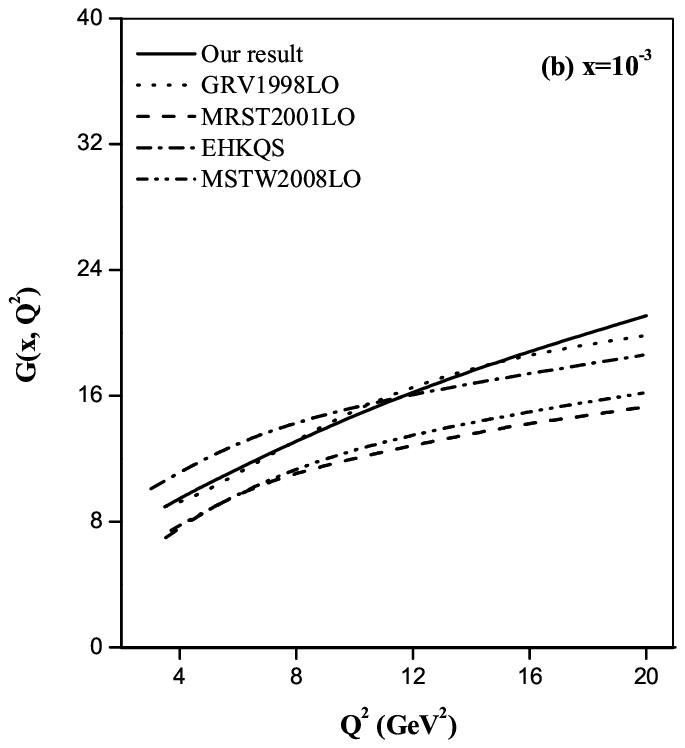}
\includegraphics[width=2.5in,height=3in]{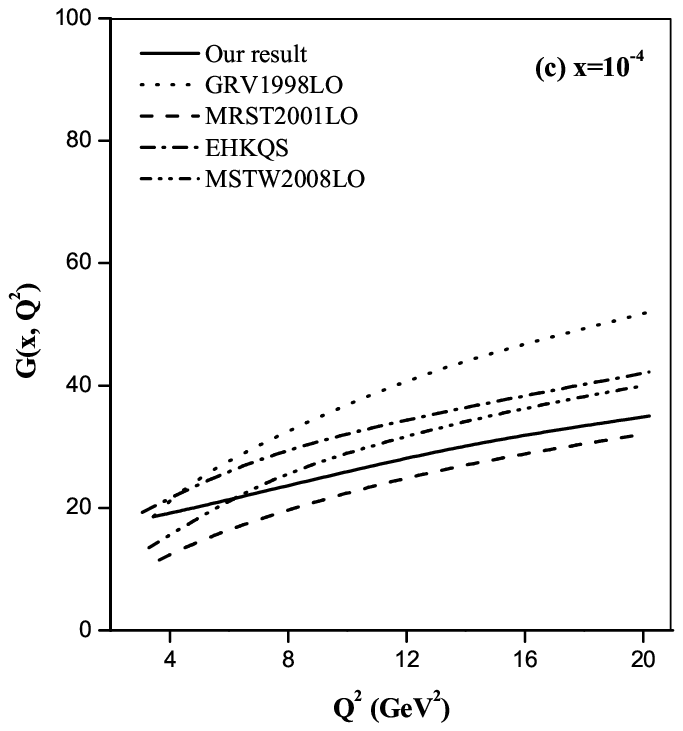}
\includegraphics[width=2.5in,height=3in]{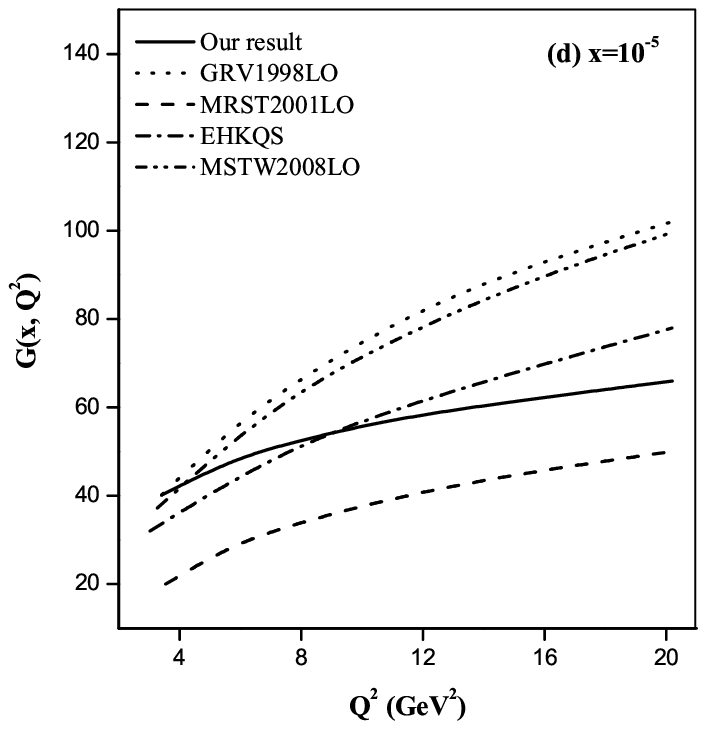}
\caption{\footnotesize {$Q^2$-evolution of $G(x,Q^2)$ for $R=2 GeV^{-1}$. Solid lines represent our results in LO whereas dotted lines are GRV1998LO results, dashed lines are MRST2001LO results, dashed dot dot lines are reuslts of MSTW2008LO and dashed dot lines are results of EHKQS model.}}
\label{fig:1}
\end{figure}

\begin{figure}[!htb]
\centering
\includegraphics[width=2.5in,height=3in]{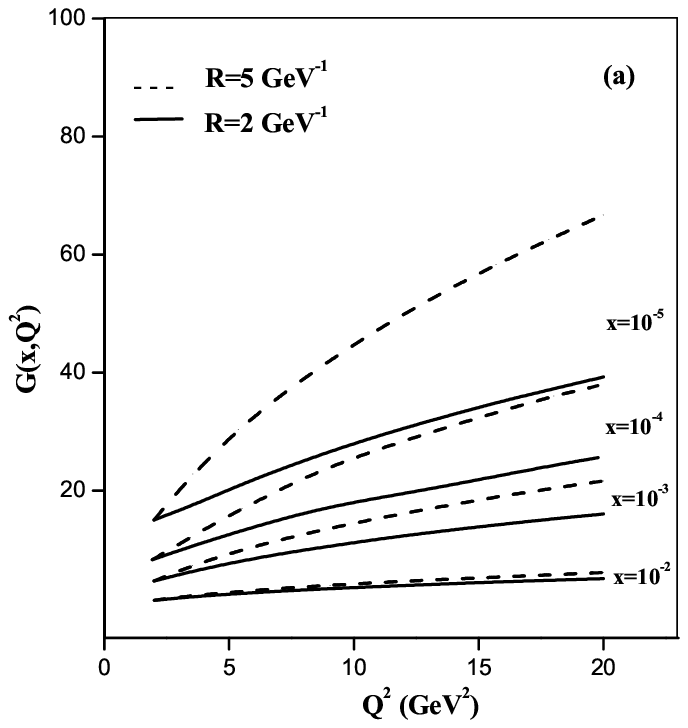}
\includegraphics[width=2.5in,height=3in]{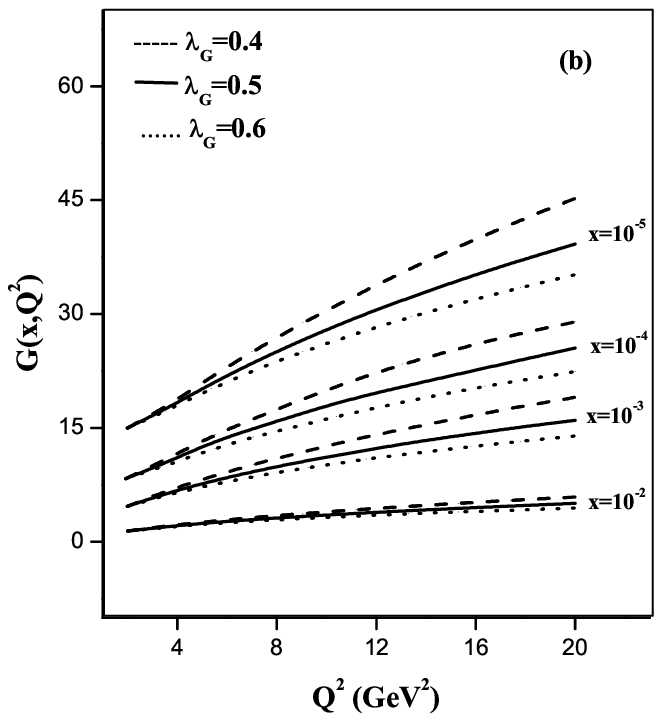}

\caption{\footnotesize {Sensitivity of $R$ and $\lambda_G$ in our results of $Q^2$-evolution of $G(x, Q^2)$.}}
\label{fig:2}
\end{figure}

 Figs. 1(a-d) represent our best fit results of $Q^2$-evolution of the gluon distribution function $G(x,Q^2)$ for $R=2\hspace{4pt}GeV^{-1}$ computed from Eq. (20) for $x=10^{-2}, 10^{-3}, 10^{-4}$ and $10^{-5}$ respectively. In all graphs the input distribution $G(x,t_0)$ at a given value of $Q_0^2$ is taken from the GRV1998LO to test the $Q^2$-evolution of $G(x,Q^2)$. In our analysis we consider the kinematic range $2\hspace{4pt}GeV^2 \leq{Q^2}\leq{20\hspace{4pt}GeV^2}$, where we expect our solution to be valid. The average value of $\Lambda$ in our calcultaion is taken to be 0.192 GeV. It is observed from the figures that our results show almost similar behaviour with those obtained from different global parametrizations and also with EHKQS model.

\ We have also investigated the effect of nonlinearity in our results for $R=2\hspace{4pt}GeV^{-1}$ and $R=5\hspace{4pt}GeV^{-1}$ respectively. For this analysis our computed values of $G(x, Q^2)$ for $R=2\hspace{4pt}GeV^{-1}$ and $R=5\hspace{4pt}GeV^{-1}$ respectively from Eq. (20) are plotted against $Q^2$ in Fig. 2(a) for $x=10^{-2}, 10^{-3}, 10^{-4}$ and $10^{-5}$ respectively. Here, the input distribution is taken from MSTW2008 global parametrization for a given value of $Q_0^2$. We have also performed an analysis to check the sensitivity of the free parameter $\lambda_G$ in our results. Fig. 2(b) represents the results for the $Q^2$-dependence of $G(x, Q^2)$ obtained from the solution of nonlinear GLR-MQ equation given by Eq. (20) for three different values of $\lambda_G$ and we observed that results are very sensitive to $\lambda_G$ as $x$ decreases.

\vspace{10pt}

\section{Conclusion}
\paragraph \ We solve the nonlinear GLR-MQ evolution equation by considering the Regge like behavior of gluon distribution function and studied the effects of adding the nonlinear GLRMQ corrections to the LO DGLAP evolution equations. Here we expect the validity of the Regge type solution of the GLR-MQ equation for gluon distribution function in the region of small $x$ and intermediate values of $Q^2$. From our phenomenological study as well we can expect our solution given by Eq. (16) to be valid in the kinematic region 2$\hspace{4pt}GeV^2 \leq Q^2 \leq 20\hspace{4pt}GeV^2$ and $10^{-5} \leq x \leq 10^{-2}$, where the nonlinear corrections cannot be neglected. Moreover, we can anticipate the Regee type solution of gluon distribution function to be valid as our obtained results of $G(x, Q^2)$ are compatible with different parameterizations. We can conclude that the solution suggested in this work given by Eq. (16) is valid only in the vicinity of saturation border. In this region one may also obtain the solution of the nonlinear equation in the form $N\propto{(Q_S(x)/q^2)^{1-\gamma_{cr}}}$, as suggested in Refs. [49 - 51]. We are also interested to obtain a solution of the nonlinear GLR-MQ equation in this form and planning to produce in a future paper.

\ We observe that the gluon distribution function increases with increasing $Q^2$ as usual which is in agreement with perturbative QCD fits at small-$x$, but with the inclusion of the nonlinear terms, $Q^2$-evolution of $G(x,Q^2)$ is slowed down relative to DGLAP gluon distribution. For the gluon distribution the nonlinear effects are found to play an increasingly important role at $x\leq10^{-3}$. The nonlinearities, however, vanish rapidly at larger values of $x$. It is also interesting to observe that nonlinearity increases with decreasing value of $R$ as expected. The differences  between the data at $R=2\hspace{4pt}GeV^{-1}$ and at $R=5\hspace{4pt}GeV^{-1}$ increase as $x$ decreases which is very clear from Fig. 2(a). Results also confirm that the steep behavior of gluon distribution function is observed at $R=5\hspace{4pt}GeV^{-1}$, whereas  it is lowered at $R=2\hspace{4pt}GeV^{-1}$ with decreasing $x$ as $Q^2$ increases. We have also investigated the sensitivity of $\lambda_G$ in our calculations and found that results are highly sensitive to $\lambda_G$ as $x$ goes on decreasing.

\section{\bf{Acknowledgement}}

\ \hspace{4pt} The authors are grateful to UGC for financial support in the form of a major research project.
\

\end{document}